\documentclass[aps,prl,twocolumn,groupedaddress,showpacs,floatfix]{revtex4}

\usepackage{graphicx}

\begin{document}

\title{Violation of the `Zero-Force Theorem' in the time-dependent
  Krieger-Li-Iafrate approximation}

\author{Michael Mundt and Stephan K\"ummel}
\affiliation{
Physikalisches Institut, Universit\"at Bayreuth, D-95440 Bayreuth,
Germany}
\author{Robert van Leeuwen}
\affiliation{University of Groningen, Theoretical Chemistry, Zernike
  Institute for Advanced Materials, 9747AG, Nijenborgh 4, Groningen,
  The Netherlands}
\author{Paul-Gerhard Reinhard}
\affiliation{Institut f\"ur Theoretische Physik II, Universit\"at
  Erlangen-N\"urnberg, Staudtstrasse 7,D-91058 Erlangen, Germany}

\begin{abstract}
We demonstrate that the time-dependent Krieger-Li-Iafrate approximation in
combination with the exchange-only functional violates the `Zero-Force
Theorem'. By analyzing the time-dependent dipole moment of Na$_5$ and Na$_9^+$,
we furthermore show that this can lead to an unphysical self-excitation of
the system depending on the system properties and the excitation strength.
Analytical aspects, especially the connection between the `Zero-Force
Theorem' and the `Generalized-Translation Invariance' of the potential, are
discussed.
\end{abstract}


\maketitle

Since the pioneering work of Ando, Peuckert, Zangwill and Soven
\cite{Ando,Peu,ZS} and the rigorous foundation by Runge and Gross
\cite{RG}, time-dependent density functional theory (TDDFT) has become
one of the most successful theories for the calculation of electronic
properties. Based on the idea of describing a system only in terms of
its time-dependent density, TDDFT offers the opportunity to access
large systems which are out of reach for any wave-function based
method. Additionally, it offers a tractable method to deal with
strong, non-linear and non-perturbative excitations of atoms,
molecules and clusters. High-harmonic generation and above-threshold
ionization are just two examples for applications of TDDFT in this
domain.

As in ground-state density functional theory, most TDDFT calculations
are done in the Kohn-Sham (KS) scheme \cite{KS}. In this approach, the
interacting many-particle system is replaced by a ficticious system of
non-interacting particles moving in a local effective potential
$v_{\mathrm{KS}}$. This potential is chosen in such a way that it
exactly reproduces the time-dependent density. As in the static case,
the potential is split into the Hartree part $v_{\mathrm{H}}$,
containing the classical electrostatic interaction, the local external
potential $v_{\mathrm{ext}}$ from the ions and other external
potentials, e.g., from a laser, and the exchange-correlation (xc)
potential $v_{\mathrm{xc}}$ containing all quantum-mechanical
exchange-correlation effects. Since the exact xc potential is not
known, it is crucial to have good approximations for it. The most
well-known approximation is the time-dependent local-density
approximation (TDLDA). It simply uses the static local-density
approximation functional for $v_{\mathrm{xc}}$ in combination with the
time-dependent density. Although being based on the ground-state
energy of the homogenous electron gas, this approximation works well
in many situations \cite{ZS,Eka84,YabBer,TDRev,PGRetal,Cast}.
Nevertheless it can also fail dramatically in some situations (see,
e.g., \cite{Petersilka,Grit,Gisb}). One of its most prominent problems
is the self-interaction error leading to a wrong asymptotic behavior
of $v_{\mathrm{xc}}$ and, as a consequence, wrong ionization
dynamics. In addition, it lacks a derivative discontinuity
\cite{Per82,Mun05} and does not include any memory effects
\cite{Maitra}.

One possible way of overcoming these problems are orbital functionals
\cite{Orbfctnal}. These functionals depend explicitly on the KS
orbitals and implicitly on the density. Thus, they are legitimate
density functionals and provide many advantages over explicit
functionals of the density. For instance, they include memory effects,
show a derivative discontinuity, and using the exact-exchange
functional (EXX) with the KS orbitals cures the Hartree
self-interaction problem. Unfortunately, they also pose a severe
problem: in order to find $v_{\mathrm{xc}}$, it is necessary to solve
the time-dependent optimized-effective potential (TDOEP) equation
\cite{TDOEP}. This is not an easy task \cite{MSOEP}. To circumvent
this problem, Ullrich {\it et al.}, already in the first publication
of the TDOEP equation, proposed an approximation to the exact
potential, namely what they called the time-dependent
Krieger-Li-Iafrate (TDKLI) approximation. TDKLI has been frequently
used to carry out TDDFT calculations with orbital functionals, e.g.,
to calculate ionization processes and high-harmonic generation
\cite{TDOEP,CAU,Chu}. In fact, besides the time-dependent common
energy denominator approximation (CEDA) \cite{CEDA,AGoer}, this is the
only practicable approximation to the exact TDOEP outside the linear
domain at present. On first sight it might be surprising that although
the EXX-TDKLI potential is not obtained as the functional derivative
of the EXX orbital functional, it is nevertheless a legitimate density
functional due to the one-to-one correspondence between the density
and the external potential applied to the non-interacting KS system.

On the fundamental side of TDDFT, many exact constraints which
$v_{\mathrm{xc}}$ must fulfill have been revealed
\cite{Hessler,RvL,UvB}. Well-known examples are the
`Harmonic-Potential theorem' \cite{DobsonHP} and the `Zero-Force
theorem' \cite{MLevy,VignaleZF,TDRev}. The statement of the latter is
that the force which the xc potential exerts on an electron cloud must
vanish, i.e.,
\begin{equation}
\int n(\mathbf{r},t) \, \nabla v_{\mathrm{xc}} (\mathbf{r},t) \ d^3r \ = \
0 \ .
\label{Eq:ZeroForce}
\end{equation}
Since the force from the Hartree potential vanishes, this guarantees
that
\begin{equation}
\partial_t \, \mathbf{P} (t) \ = \ - \int n(\mathbf{r},t) \, \nabla
v_{\mathrm{ext}} (\mathbf{r},t) \ d^3r
\label{Eq:DtPexact}
\end{equation}
holds with $\mathbf{P} (t)$ being the electronic momentum of the
system \cite{RvL}. From earlier work it is known that the EXX
functional is a `conserving approximation' \cite{UvB} {\it if} the
full TDOEP equation is solved. However, since this is a crucial
requirement for the proof, the situation is less clear for the case of
the EXX-TDKLI potential. It is the aim of this manuscript to explore
whether the EXX-TDKLI nevertheless satisfies the `Zero-Force
theorem'.

In order to answer this question, we have calculated the response of a
Na$_5$ cluster to a small dipole excitation. This cluster has planar
geometry (oriented here in the $(x-y)$-plane) and a rather ``soft''
electron cloud which provides a critical test case for our
purposes. For the calculation we used a modified version of the PARSEC
program \cite{Parsec} in combination with a local pseudopotential for
the Na cores \cite{SKPP}. For the real-space grid a sphere of radius
20 $a_0$ and a grid spacing of 0.7 $a_0$ was used. The kinetic energy
part of the KS Hamiltonian was approximated by a 12-point formula for
the Laplacian. After the ground-state calculation, we applied a
momentum boost $\exp (i \, \mathbf{r} \cdot
\mathbf{p}_{\mathrm{boost}} / \hbar)$ with
$|\mathrm{p}_{\mathrm{boost}}| = 3.834 \times 10^{-4} \, \hbar
a_0^{-1}$, corresponding to a total excitation energy of the system of
$1.0 \times 10^{-5}$ eV, to all KS orbitals. The boost had an equal
strength in $x$-, $y$-, and $z$-direction. The resulting excited state
was propagated in real time with fixed ions. The propagation was done
with a fourth-order Taylor approximation to the propagator. We checked
the convergence of our results with respect to the time step and found
a value of $\Delta t = 0.003$ fs for the propagation sufficient.

\begin{figure}[!t]
\includegraphics[width=8.5cm]{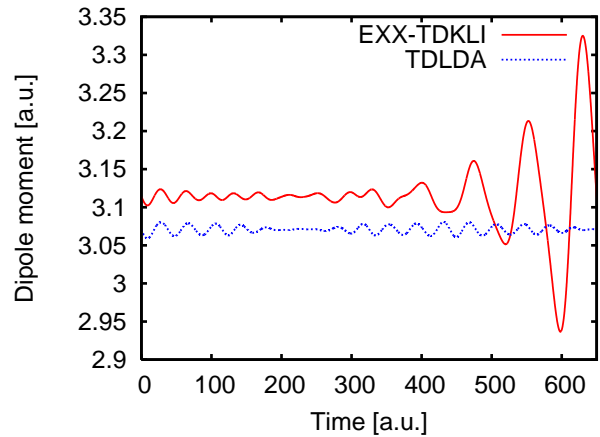}
\caption{\label{Fig:1} (Color online) $y$-component of the dipole moment of
  Na$_5$ after an initial boost of $|\mathrm{p}_{\mathrm{boost}}| = 3.834
  \times 10^{-4} \, \hbar a_0^{-1}$. In contrast to the TDLDA result, the
  EXX-TDKLI curve shows an increasing amplitude of the oscillation. The
  EXX-TDKLI curve is slightly shifted for better comparison. Rydberg atomic
  units are used throughout.}
\end{figure}

Fig.\ \ref{Fig:1} shows the resulting $y$-component of the dipole
moment $d_y(t) = e \int y \, n(\mathbf{r},t) \ d^3r$ (with $e$ being
the electron's charge).
For a while the amplitude shows reasonable oscillations in agreement
with the initial boost. But after about 400 a.u., it increases rapidly
and steadily. For comparison, the same quantity is plotted for a TDLDA
calculation. There, no increasing amplitude is observed. Since the
TDLDA satisfies the `Zero-Force theorem' \cite{VignaleZF}, this is
already a hint that the increasing amplitude in the TDKLI calculation
is related to a violation of the `Zero-Force theorem'.

To demonstrate that this is indeed the case, we have monitored the
expected time-derivative of the total momentum, i.e., we have
calculated the right-hand side of Eq.\ (\ref{Eq:DtPexact}) with the
external potential coming from the local pseudopotentials of the ions
\cite{SKPP}.

\begin{figure}[!t]
\includegraphics[width=8.5cm]{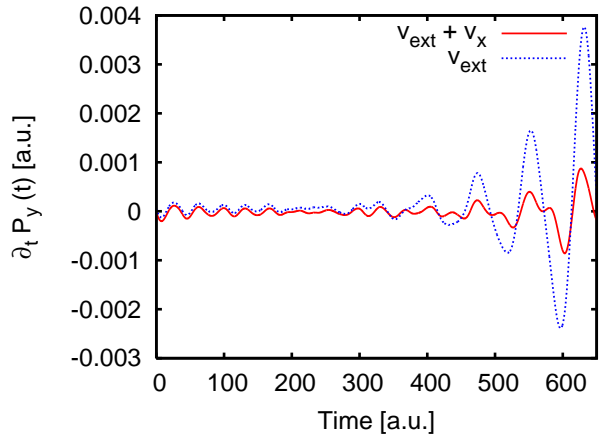}
\caption{\label{Fig:2} (Color online) Time derivative of the $y$-component of
  the total momentum of the electron density. `$v_{\mathrm{ext}} + v_{\mathrm
  x}$' and `$v_{\mathrm{ext}}$' label the results calculated via Eq.\
  (\ref{Eq:DtPexact}) with the external potential or, respectively, the
  external and the exchange potential. The violation of the `Zero-Force
  theorem' is visible in the difference between `$v_{\mathrm{ext}} +
  v_{\mathrm x}$' and `$v_{\mathrm{ext}}$'.}
\end{figure}

In addition, we have evaluated the left-hand side of Eq.\
(\ref{Eq:ZeroForce}), which should be zero. Fig.\ \ref{Fig:2} shows
the result obtained from Eq.\ (\ref{Eq:DtPexact}) and the sum of Eq.\
(\ref{Eq:DtPexact}) and Eq.\ (\ref{Eq:ZeroForce}).
One observes that the total force from the external potential on the
electron density, Eq.\ (\ref{Eq:DtPexact}), differs significantly from
the total force obtained from the sum of both equations. This clearly
demonstrates that $v_{\mathrm x}$ violates the `Zero-Force theorem'
and contributes to the total force. Since the time-dependent dipole
moment is connected to the total force $\mathbf{F}(t)$ via (see, e.g.,
\cite{RvL})
\begin{equation}
\partial_t^2 \mathbf{d}(t) \ = \ e \, \partial_t \mathbf{P}(t) \ = \
e \, \mathbf{F}(t) \ ,
\label{Eq:ForceMomenDipole}
\end{equation}
it follows that the violation of the `Zero-Force theorem' leads to a
wrong time-dependent dipole moment and to a self-excitation of the
system. Finally, we have checked that Eq.\ (\ref{Eq:ForceMomenDipole})
holds for the KS system. For this we calculated the force from the
total KS potential, i.e.,
\begin{equation}
\mathbf{F}(t) \ := \ - \int n(\mathbf{r},t) \, \nabla v_{\mathrm{KS}}
(\mathbf{r},t) \ d^3r \ ,
\end{equation}
and compared it to the time-derivative of the KS current
\begin{equation}
\mathbf{j}_{\mathrm{KS}} (\mathbf{r},t) \ = \ \frac{\hbar^2}{2 m i}
\sum_{k=1}^{N} (\varphi_k^*(\mathbf{r},t) \nabla \varphi_k(\mathbf{r},t) \, -
\, c.c.)
\end{equation}
from our calculation. The current is connected to the total momentum via
$\mathbf{P}(t) = m \int \mathbf{j}_{\mathrm{KS}} (\mathbf{r},t) \ d^3r$. By
explicitly calculating the second time-derivative of the monitored dipole
signal, we confirmed that Eq.\ (\ref{Eq:ForceMomenDipole}) holds in our
calculation as it should be.

\begin{figure}[!t]
\includegraphics[width=8.5cm]{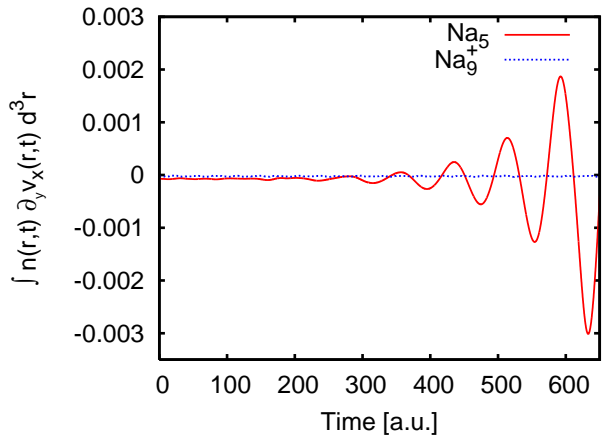}
\caption{\label{Fig:3} (Color online) Violation of the `Zero-Force theorem'
  for Na$_9^+$ and Na$_5$. Again, the $y$-component is plotted. Both systems
  were excited by a momentum boost corresponding to an excitation of the
  system by $1.0 \times 10^{-5}$ eV. In contrast to Na$_5$, the resulting
  curve for Na$_9^+$ does not increase in time. The resulting dipole moment of
  Na$_9^+$ is also observed to be stable.}
\end{figure}

To study the influence of the `Zero-Force theorem' violation in more detail, we
have carried out the same calculation for a Na$_9^+$ cluster. In contrast to
Na$_5$, this cluster can be considered as one of the most `forgiving' systems
because of its spherical shape and the positive charge leading to a stable
`plasmon'-like oscillation when excited. Fig.\ \ref{Fig:3} shows the resulting
violation of Eq.\ (\ref{Eq:ZeroForce}).
A close look reveals that the `Zero-Force theorem'
is slightly violated again. But, in contrast to Na$_5$, the violation
does not increase in time. Checking the time-dependent dipole moment
shows that it is also stable. Beside the stronger binding forces a
possible explanation for this observation could be that the higher
inversion symmetry of the almost spherical Na$_9^+$ leads to an error
cancellation in the course of one density oscillation and thus to a
strongly reduced increase of the violation. In any case, the result
clearly corroborates the intuitive expectation that the system
properties have a strong influence on the degree of the violation
of the `Zero-Force theorem'.

\begin{figure}[!t]
\includegraphics[width=8.5cm]{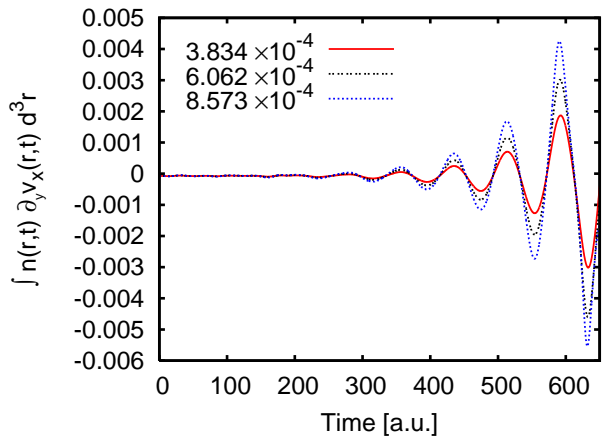}
\caption{\label{Fig:4} (Color online) Violation of the `Zero-Force theorem'
  for Na$_5$ excited with different boosts of $3.834 \times 10^{-4} \, \hbar
  a_0^{-1}$, $6.062 \times 10^{-4} \, \hbar a_0^{-1}$, and $8.573 \times
  10^{-4} \, \hbar a_0^{-1}$. Higher excitation energies clearly lead to an
  increased violation. As in Fig.\ \ref{Fig:2} only the $y$-component is
  plotted.}
\end{figure}

In addition to the system properties, one can also expect an influence
of the excitation energy on the `Zero-Force' violation. And indeed,
this can be found. Fig.\ \ref{Fig:4} shows the results for Na$_5$
excited with three different boost strengths of $3.834 \times 10^{-4}
\, \hbar a_0^{-1}$, $6.062 \times 10^{-4} \, \hbar a_0^{-1}$, and
$8.573 \times 10^{-4} \, \hbar a_0^{-1}$.
Obviously, the deviation from zero varies with the boost strength. The
importance of the excitation energy can also be demonstrated by
considering the extreme situation of no excitation at all. In this
case, the time-dependent KS orbitals are given by the ground-state
orbitals multiplied by a time-dependent phase factor $\exp (-i
\epsilon_k t / \hbar)$ containing the KS eigenvalue $\epsilon_k$. But
since this phase factor does not influence the potential, the whole
system remains in a stationary state and the violation of the
`Zero-Force theorem' remains constant. This constant violation does
not lead to a non-stationary state since $\int n(\mathbf{r},t) \,
\nabla v_{\mathrm{KS}} (\mathbf{r},t) \ d^3r$ vanishes due to the
ground-state iteration. In other words, for the ground-state density
the total force from the external and the xc potential are in
equilibrium leading to a stationary state.

Beside the just discussed aspects, there are three other conditions under
which a violation of the `Zero-Force theorem' in a TDKLI calculation
may not be observed. First, for short time scales the accumulation of
the violation can be too small to show up significantly. Certainly,
this time scale depends on the two aspects discussed above: the
excitation strength and the system properties. The second situation
occurs when a strong, ionizing external field is applied to the
system. This can hide the error in the xc potential
completely. Finally, for spin-saturated two-particle systems the TDKLI
and the TDOEP potential coincide and, as a consequence, the
`Zero-Force theorem' is rigorously satisfied if the TDKLI potential
comes from a `conserving approximation' \cite{UvB,BayKad}. Since, to
the best of our knowledge, one of these three conditions can be found
in all applications of the TDKLI potential to date, it is not
surprising that the violation of the `Zero-Force theorem' was not
reported earlier.

After the numerical results we now focus on analytical
considerations. As Vignale has shown, any potential obtained from a
`generalized-translation invariant' xc action functional satisfies the
`Zero-Force theorem' \cite{VignaleZF,UvB}. In addition, the potential
also satisfies `Generalized-Translation invariance', i.e., $v_{\mathrm
  xc}$ rigidly follows a rigidly translated density, and, as a
consequence, the `Harmonic-Potential theorem'. Actually, the same
arguments show that if $v_{\mathrm xc}$ satisfies
`Generalized-Translation invariance' {\it and} is the functional
derivative of some xc action functional, it must also satisfy the
`Zero-Force theorem'. Since the EXX-TDKLI satisfies the
`Harmonic-Potential theorem' \cite{TDRev,Remark}, we conclude from our
numerical results that the EXX-TDKLI potential in general cannot be
obtained as the functional derivative of some xc action functional
with respect to the density. This result is in line with earlier
results for the static Slater potential \cite{MLevyPRL}. The missing
action functional for the EXX-TDKLI potential, in combination with the
observation that the TDKLI potential is nevertheless used as part of a
legitimate density functional procedure, makes any rigorous analytical
statements about the properties of the potential extremely
difficult. A hand-waving argument, however, is provided by the
expression for the EXX-TDOEP involving the orbitals and orbital shifts
\cite{TDOEP,MSOEP}. Since the full expression satisfies the
`Zero-Force theorem', it is not surprising that the TDKLI
approximation, which neglects the terms containing the orbital shifts,
violates the `Zero-Force theorem'. Additionally, the violation of the
`Zero-Force theorem' in the static Slater approximation
\cite{MLevyPRL} supports the presented numerical results, too.

Finally, we want to comment on the generalization of our results to
TDKLI potentials from other xc orbital functionals for the
action. Clearly, the possibility that other expressions lead to a
TDKLI potential which satisfies the `Zero-Force theorem' cannot be
ruled out from our calculations. However, since the EXX functional is
a `conserving approximation' \cite{UvB} and this property is destroyed
by using the TDKLI and not the TDOEP potential, it is highly plausible
that this also happens if other `conserving approximations'
\cite{UvB,BayKad} are used. In any case, it is highly recommendable to
check the `Zero-Force theorem' in any TDKLI calculation.

To summarize, we have demonstrated that the TDKLI approximation in
combination with the exact-exchange functional violates the
`Zero-Force theorem'. By calculating the response of Na$_5$ and
Na$_9^+$, we have demonstrated that this violation can lead to a
self-excitation of the system and a wrong time-dependent dipole
moment. Furthermore, a strong system and excitation-energy dependence
has been observed. Due to the difficulties associated with the
construction of orbital-dependent approximations for the xc potential
which satisfy the `Zero-Force theorem', our findings clearly
demonstrate the urgent need for a working TDOEP scheme in order to
make orbital functionals accessible for real-time TDDFT.

\begin{acknowledgments}
S.\ K.\ thanks the Deutsche Forschungsgemeinschaft for financial support. R.\
v.\ L.\ thanks the Dutch Foundation of Fundamental Research of Matter (FOM)
for support.
\end{acknowledgments}


\begin{thebibliography}{100}
\bibitem{Remark}As a consequence, the EXX-TDKLI potential does not
  lead to self-excitation in the case of systems confined by a
  harmonic external potential and a perturbing spatially homogeneous
  time-dependent force. In order to test our numerical implementation
  we have checked this and can confirm that the left-hand side of Eq.\
  (\ref{Eq:ZeroForce}) remains constant.
\end{thebibliography}
\end{document}